\documentclass{article}

\usepackage{PRIMEarxiv}

\usepackage[utf8]{inputenc} 
\usepackage[T1]{fontenc}    
\usepackage{hyperref}       
\usepackage{url}            
\usepackage{booktabs}       
\usepackage{amsfonts}       
\usepackage{nicefrac}       
\usepackage{microtype}      
\usepackage{lipsum}
\usepackage{fancyhdr}       
\usepackage{graphicx}       
\graphicspath{{media/}}     
\usepackage{amsmath}

\title{Reduced Order Dynamical Models for Complex Dynamics in Manufacturing and Natural Systems Using Machine Learning 
\thanks{\textit{\underline{Citation}}: 
\textbf{Authors. Title. Pages.... DOI:000000/11111.}} 
}

\author{
  William Farlessyost\\
  Agricultural \& Biological Engineering \\
  Purdue University \\
  West Lafayette, IN, USA\\
  \texttt{wfarless@purdue.edu} \\
   \And
  Shweta Singh\\
  Agricultural \& Biological Engineering \\
  Environmental \& Ecological Engineering\\
  Purdue University \\
  West Lafayette, IN, USA\\
  \texttt{singh294@purdue.edu} \\
}

\begin{document}
\maketitle

\begin{abstract}
Dynamical analysis of manufacturing and natural systems provides critical information about production of manufactured and natural resources respectively, thus playing an important role in assessing sustainability of these systems. However, current dynamic models for these systems exist as mechanistic models, simulation of which is computationally intensive and does not provide a simplified understanding of the mechanisms driving the overall dynamics. For such systems,  lower-order models can prove useful to enable sustainability analysis through coupled dynamical analysis. There have been few attempts at finding low-order models of manufacturing and natural systems, with existing work focused on model development of individual mechanism level. This work seeks to fill this current gap in the literature of developing simplified dynamical models for these systems by developing reduced-order models using a machine learning (ML) approach. The approach is demonstrated on an entire soybean-oil to soybean-diesel process plant and a lake system. We use a grey-box ML method with a standard nonlinear optimization approach to identify relevant models of governing dynamics as ODEs using the data simulated from mechanistic models. Results show that the method identifies a high accuracy linear ODE models for the process plant, reflective of underlying linear stoichiometric mechanisms and mass balance driving the dynamics. For the natural systems, we modify the ML approach to include the effect of past dynamics, which gives  non-linear ODE. While the modified approach provides a better match to dynamics of stream flow, it falls short of completely recreating the dynamics. We conclude that the proposed ML approach work well for systems where dynamics is smooth, such as in manufacturing plant whereas does not work perfectly well in case of chaotic dynamics such as water stream flow.

\end{abstract}


\section{Introduction}
Process industries can be defined as those which apply chemical or mechanical changes to their system inputs to output a product in a continuous or semi-continuous fashion ~\cite{williams1963process}. System identification of these processes is crucial within their respective industries for developing models that can be used for plant design, observation or control. This system identification is typically iterative and data-driven since \textit{a priori} model structure is typically minimal ~\cite{rivera2003plant}. To limit the amount of disturbance to plant operation, these system identification methods must be ``plant-friendly," meaning industries go to great lengths to use data collection experiments that minimizes equipment degradation, plant output deviations, and experiment time ~\cite{rivera2003plant}. As a result, the recovered dynamical models from these efforts are rarely published or made publicly available by industry.

This lack of model availability limits research external to industry that requires model-based understanding of industrial process dynamics. While highly accurate mechanistic models based on first-principle relationships exist and are used in chemical process dynamics modeling software like Aspen Plus\textsuperscript{TM}, the prohibitively high order of these models limits their interpretation as white-box models, in applications requiring computational simplicity, and as component models that may be coupled to establish a model for a larger system. 

One area of research limited by dynamic model availability is sustainability analysis of industrial processes within a larger industrial and ecological network. Here low-order dynamic models that capture phase space relationships of critical mass and energy flows between industries or ecological processes would provide researchers with insight into both long and short-term trajectories of these coupled systems. Additionally, reduced-order models that can be quickly numerically integrated for fast simulation might be used to approximate the answer to a number of ``What-If?" scenarios for specific inputs and initial conditions.

We propose that new system identification methods employing sparse regression can be used for recovering simplified models for a variety of different complex systems, including process plants and watershed mechanics. While a number of variations have arisen since the initial publication of \cite{brunton2016sparse}, the core method, the sparse identification of nonlinear dynamics (SINDy) algorithm, provides a method of system identification that makes minimal assumptions about the physics of the system or the necessary model structure. Rather, the model structure is assumed to be a sum of potential functions, with coefficients that are regressed with intermittent thresholding in order to maximize accuracy and ensure sparsity \cite{brunton2016discovering}.

In \cite{brunton2016sparse}, the SINDy algorithm outlined in \cite{brunton2016discovering} is expanded to include methods for handling control and forcing inputs to a system.The authors show the performance of this algorithm on both a Lotka-Volterra predator-prey model and the Lorenz System with external inputs to simultaneously demonstrate the failure of naive SINDy, with no consideration of input, and the success of the modified SINDy algorithm when these inputs are taken into account. The authors of \cite{stender2019recovery} suggest that the SINDy algorithm described in \cite{brunton2016discovering} and \cite{brunton2016sparse} may be further improved by employing bounded nonlinear optimization of the model coefficients. We apply both the input and optimization modification in our implementation of SINDy. 

In \cite{subramanian2021white}, SINDy is applied and tested as a method for learning a dynamic model of a distillation column. This method is tested along with symbolic regression, and the two are compared on overall accuracy and model composition. The authors of \cite{subramanian2021white} find that SINDy recovers models of higher complexity that outperform the symbolic regression, while symbolic regression recovers simpler models that outperform the SINDy method over longer intervals. The success here in using SINDy to learn a model for a dynamic process as complex as a distillation column, a single unit operation, seems to suggest that SINDy might also be effective when applied to a collection of unit operations comprising an entire process plant. 

 We employ the SINDy algorithm to recover a dynamical model for a soybean-oil to soybean-diesel transesterification process using simulated time-series data. In particular, we model the dynamic behavior of material flow rates at various points in the process. While publicly available models do exist for the kinetics of soybean-oil transesterification and the dynamics within the plant reactor ~\cite{diasakou1998kinetics, wenzel2006modelling, zapata2018different}, plant-wide models that also capture the dynamic relationship between the internal molar flow rates, the output soybean-diesel, and the input flows is not available. This makes the soybean-oil to soybean-diesel process a useful test case for this system identification method. 
 
 We further apply the SINDy algorithm in an attempt to recover a low-level dynamic equation for streamflow dynamics of the North Fork Vermilion River, providing water to the town of Danville, Illinois, by training on historical streamflow and climate data. The town of Danville contains one of the largest soybean-oil to soybean-diesel plants in Illinois, so finding such an equation provides a method of directly analyzing the long term water flow under different forcing functions. These simplified dynamical equations for the natural water system can also be used to couple these systems with other dynamical systems. 
 

 The remaining paper structure is organized as follows. Section 2 covers the methodology of the SINDy algorithm, selection of relevant state variables, and data generation or selection. Section 3 describes the model recovery results and Section 4 follows with relevant conclusions and discussion.

\section{Materials and Methods}

\subsection{SINDy Algorithm} 
The SINDy method, as described in \cite{brunton2016discovering}, assumes the system in question can be modeled using ordinary differential equation type state equations of the form

\begin{equation}
    \mathbf{\dot x}(t) = \mathbf{f}(\mathbf{x} (t)),
\end{equation}

\noindent where $\mathbf{x}(t) \in \mathbf{R}^n$ is a vector of state variables at time $t$, and $\mathbf{f}(\mathbf{x} (t))$ are the equations defining the dynamics of the system. To determine an optimal model structure and parameterization for the function, $\mathbf{f}$, we begin by collecting time-series data for the system states, $\mathbf{x} (t)$ sampled at times $t_1, t_2, ... , t_m$. This can be arranged in a matrix, $\mathbf{X}$, as 

\begin{equation}
    \mathbf{X} =    \begin{bmatrix} \mathbf{x}^T (t_1) \\   \mathbf{x}^T (t_2) \\    \vdots                   \\    \mathbf{x}^T (t_m)    \end{bmatrix}   = \begin{bmatrix}   x_1 (t_1) & x_2 (t_1) & \hdots & x_n (t_1) \\    x_1 (t_2) & x_2 (t_2) & \hdots & x_n (t_2) \\    \vdots & \vdots & \ddots & \vdots                   \\    x_1 (t_m) & \mathbf{x_2} (t_m) & \hdots & x_n (t_m)    \end{bmatrix}.
\end{equation}

\noindent We then numerically determine the time derivative of these states, $\mathbf{\dot x} (t)$, and arrange it in a similar matrix, $\mathbf{\dot X}$, as 

\begin{equation}    \mathbf{\dot X} =    \begin{bmatrix}    \mathbf{\dot x}^T (t_1) \\    \mathbf{\dot x}^T (t_2) \\    \vdots                   \\    \mathbf{\dot x}^T (t_m)    \end{bmatrix}   = \begin{bmatrix}    \dot x_1 (t_1) & \dot x_2 (t_1) & \hdots & \dot x_n (t_1) \\    \dot x_1 (t_2) & \dot x_2 (t_2) & \hdots & \dot x_n (t_2) \\    \vdots & \vdots & \ddots & \vdots                   \\    \dot x_1 (t_m) & \dot x_2 (t_m) & \hdots & \dot x_n (t_m)    \end{bmatrix}.
\end{equation}

\noindent While we know the model will be composed of a sum of different component functions, we do not know which functions the sparse regression algorithm will select. Therefore, we provide a library of candidate functions, $\mathbf \Theta (\mathbf{X})$, in the form

\begin{equation}
\mathbf \Theta (\mathbf{X}) = \begin{bmatrix}    \mathbf{1} & \mathbf{X} & \mathbf{X}^{P_2} & \hdots & \sin(\mathbf{X}) & e^{\mathbf{X}} & \hdots     \end{bmatrix}
\end{equation}

\noindent where $\mathbf{X}^{P_2}$ are possible quadratic nonlinearities in $\mathbf x$, $\mathbf{X}^{P_3}$ are possible cubic nonlinearities, and so on. For example, 

\begin{equation}
  \mathbf{X}^{P_2} = \begin{bmatrix}    x_1^2 (t_1) & \omega (t_1) & x_2^2 (t_1) & \hdots & x_n^2 (t_1) \\    x_1^2 (t_2) & \omega (t_2) & x_2^2 (t_2) & \hdots & x_n^2 (t_2) \\    \vdots & \vdots & \ddots & \vdots    \\    x_1^2 (t_m) & \omega (t_m) & x_2^2 (t_m) & \hdots & x_n^2 (t_m)     \end{bmatrix}
\end{equation}

\noindent where $\omega (t) = x_1 (t) x_2 (t)$ for compactness.

\noindent We determine which functions in $\mathbf \Theta (\mathbf{X})$ will be included in the model by solving the sparse regression problem given by 

\begin{equation}
    \mathbf{\dot X} = \mathbf \Theta (\mathbf{X}) \mathbf{\Xi}
\end{equation}

\noindent where 

\begin{equation}
\mathbf{\Xi} = \begin{bmatrix} \boldsymbol{\xi}_1 & \boldsymbol{\xi}_2 & \hdots & \boldsymbol{\xi}_n \end{bmatrix}
\end{equation}

\noindent is a matrix of sparse vectors of coefficients. When data is collected from real world experiments or is known to be noisy, an additional $\mathbf{Z}$ matrix can be added to the right side of the sparse regression problem to account for this noise. However, we exclude this term since our simulation data is known to not contain noise, owing to the functioning of the ASPEN Plus Dynamics simulation.

After solving $\mathbf{\Xi}$, the model can be written as 

\begin{equation}
    \mathbf{\dot x}_k = \mathbf \Theta (\mathbf{x}^T) \boldsymbol{\xi}_k
\end{equation}

\noindent for every row $k$ of the state equations. 

To account for some set of input signals,  $\mathbf{u} (t)$, driving the system, we assume the system can instead be modeled using state equations of the form

\begin{equation}
    \mathbf{\dot x}(t) = \mathbf{f}(\mathbf{x} (t),\mathbf{u} (t)),
\end{equation}
\noindent as shown by \cite{brunton2016sparse}. The SINDy algorithm remains unchanged with the regression problem now written as 

\begin{equation}
    \mathbf{\dot X} = \mathbf \Theta (\mathbf{X},\mathbf{U}) \mathbf{\Xi}.
\end{equation}
\subsection{SINDy Model Improvement: Further Nonlinear Optimization}
To improve the performance of our SINDy-recovered models, we further optimize the associated sparsity matrix, $\Xi$, by applying a constrained nonlinear optimization scheme. This optimization beyond the SINDy method is based on the work of \cite{stender2019recovery}, which suggests that while SINDy is capable of finding the location of nonzero elements of $\Xi$, it cannot necessarily find optimal values for each since $\Xi$ is discontinuous over $\lambda$ \cite{stender2019recovery}. We apply the method outlined in \cite{stender2019recovery} of sequential quadratic programming (SQP) implemented using MATLAB's \textit{fmincon}. Here we set upper and lower bounds for each nonzero element of $\Xi$ as the given constraints to \textit{fmincon} and construct an optimization function using the mean absolute error (MAE) across all state variables between the training data and the integrated model.

\subsection{SINDy Model Improvement: Inclusion of Input Derivatives}
We consider the time-derivative of each input as an additional input to the system to try and account for hysteresis when modeling streamflow using the SINDy method. That is to say, we now assume that the system can be represented as a state equation in the form of 

\begin{equation}
    \mathbf{\dot x}(t) = \mathbf{f}(\mathbf{x} (t),\mathbf{u} (t),\mathbf{\dot u} (t)),
\end{equation}

\noindent where the SINDy regression problem becomes 

\begin{equation}
    \mathbf{\dot X} = \mathbf \Theta (\mathbf{X},\mathbf{U},\mathbf{\dot U}) \mathbf{\Xi}.
\end{equation}
\noindent This additional input provides a way to tie the current state of the system to past inputs that otherwise would have no way to assert influence on the system trajectory in the model structure. It should be noted, however, that any noise in the measured input signals will propagate to the input derivative and may negatively impact recovered models. 

\subsection{Evaluation Criteria for Selection of Trained Models}

\subsubsection{Evaluation Criteria of Soybean-Diesel Plant Models}

To make an initial determination of model performance that would allow us to select a set of ``successful" models, we divide the first 150 hours of the process simulation data into five distinct folds, and implement a 5-fold cross-validation training scheme. This results in five different models recovered across the training data. We vary the value of the sparsity parameter $\lambda$ and use this 5-fold cross-validation scheme on each value. Each model is integrated over time and the mean absolute error (MAE) is computed using equation \ref{MAEequ}.
\begin{equation}
\label{MAEequ}
    MAE = \frac{ \sum | x_{r,i} - x_{m,i} |}{n},
\end{equation}

\noindent In equation \ref{MAEequ}, $x_{r}$ and $x_m$ are the state variable measurements from the test data and the model estimations respectively for each time index $i$, and $n$ is the size of the test data, between the integrated model and the simulation data from the fold. The model with the lowest MAE of these five models is selected as the representative of this value of $\lambda$.

Of these selected models, we make an additional selection based on the lowest MAE value. These high-performing models are then tested on the remaining 50 hours of data from the same simulation source as the training data. We also test the model for long term accuracy and stability by testing on 200 hours of simulation data driven by inputs with different random seeds than those in the training, validation, and 50 hour test set.

\subsubsection{Evaluation Criteria of Streamflow Models}

Due to the limited number of data points available for training and testing of the SINDy recovered streamflow models, we use the 5-fold validation technique previously described for evaluation of the soybean-diesel plant model as our sole evaluation criteria. Whereas for the soybean-diesel plant we only consider first order polynomials in the SINDy function library, we expand our search to include second order polynomials in the streamflow models. Subsequently, we vary the model degree as well as the sparsity parameter and utilize the 5-fold cross-validation scheme with MAE (equation \ref{MAEequ}) as the error metric in order to reduce the space of possible models.

\subsection{System Selection and Data Collection}

\subsubsection{Industrial System: Soybean-Oil to Soybean-Diesel Plant}
\begin{figure}[h!]
    \centering
    \includegraphics[scale=.45]{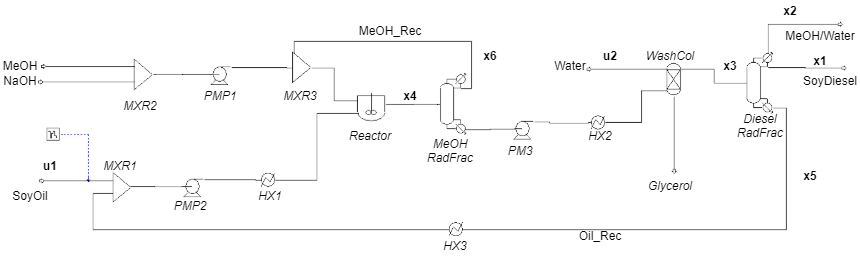}
    \caption{Soybean-oil to soybean-diesel process with state variables labeled in bold.}
    \label{fig:process_model}
\end{figure}

As shown in Figure~\ref{fig:process_model}, soybean-oil is converted to a soybean-diesel product using a series of transesterification reactions in the presence of sodium hydroxide (NaOH) mixed with methanol (MeOH). The chemical content of this soybean-oil is provided in \cite{makeri2016comparative}. The soybean-oil undergoes a transesterification process in a continuous stirred-tank reactor (CSTR). This reaction produces a mixture of methylated fatty acid molecules, glycerol, unreacted intermediate products, NaOH, and MeOH. The remaining MeOH is then separated from the other components using a RadFrac separation column and reused. A wash column removes the glycerol from the remaining mixture. The soybean-diesel and unreacted intermediate products pass through another RadFrac separation column to separate these two components. The unreacted intermediate products are then mixed with the input stream of soybean-oil to repeat the transesterification process until fully converted.

To simplify our model, we fix a number of system parameters to be constant values including: 

\begin{itemize}
\item   MeOH molar flowrate, temperature and pressure;
\item	NaOH molar flowrate, temperature and pressure;
\item	Pressure difference of pumps:
\item   Duty of heat exchangers;
\item   Vessel geometry of Reactor, MeOH and Diesel RadFrac blocks;
\item   RadFrac block stage pressures.
\end{itemize}

\noindent \textbf{State Variable Selection for Soybean-Diesel Plant :}
Selection of state variables to include in a model is one of the most important elements of system identification. Failure to select relevant variables will lead to poor model performance, no matter the tuning of model parameters. In traditional system identification, model structure is typically derived from some understanding of the underlying physics of system. However, with a complex dynamic system, such as the chemical process plant considered here, drawing from any first principle understanding may be useless due to the sheer number of variables. 

Since our primary objective is to achieve a model structure relating the soybean-diesel output to the soybean-oil and water input as molar flow rates, $x_1$, $u_1$, and $u_2$ in Figure \ref{fig:process_model} respectively, we select similar molar flow rate state variables likely consisting of relevant dynamics and operating on similar time-scales. We choose the state variables marked in Figure \ref{fig:process_model} as $x_1,...,x_6$ to be those considered in the model. Other molar flow rates are assumed to likely be redundant and are therefore unnecessary to include. Additionally, the glycerol output in Figure \ref{fig:process_model} is excluded as preliminary results indicated that the inclusion of this state variable greatly reduced model performance, likely due to nonlinearities in the wash column or dependencies on other variables not considered in the model.


\noindent \textbf{Soybean-Diesel Plant: System Excitation and Data Collection }
Based on the work of \cite{barker2003performance} we assume that for a system composed of dynamic linearities and static nonlinearities that can be expressed as a block structure, there exists an optimal pseudo-random multilevel sequence generated from Galois field polynomials that is sufficiently exciting. The authors of \cite{barker2003performance} here define optimal to mean the minimal number of levels and mention that levels beyond this minimum level do not necessarily perform better. Therefore, it is assumed that levels beyond the optimal level are unnecessary but not detrimental to system identification. This last point allows us to make use of the pseudo-random binary sequence (PRBS) block in ASPEN Plus dynamics with random amplitude for excitation of the system. By setting the amplitude to variable, the signal switches from a two-level PRBS to a sequence from Galois Field polynomials with an arbitrarily high level. Amplitude bounds and period are then varied until the state variable response appears to oscillate and not simply decrease or increase in the long term. Unfortunately, while this visual observation and adjustment cycle is a crude adjustment strategy for determining an amplitude and frequency that is sufficiently exciting, it fails to detect which of these might drive the system outside of standard operating bounds, or into other dynamic regimes entirely. The system is then simulated for 200 hundred hours and the state variable values are measured every 0.02 hours. \\

\subsubsection{Natural System: Lake Vermilion Water Supply}
Lake Vermilion, located in Danville, Illinois, is fed by the North Fork Vermilion River watershed: an approximately 295 square mile situated in Vermilion and Iroquois Counties, Illinois as well as Warren and Benton Counties, Indiana \cite{johnston2008watershed}. The lake was formed by damning the North Fork Vermilion River in 1925 and currently holds around three billion gallons of water after the lake level was raised in 1991 due to projected population increase as well as sedimentation \cite{johnston2008watershed}. However, sedimentation is estimated to continue at a rate that will reduce the lake water storage capacity by around one-percent per year \cite{bogner1999sedimentation}.

As of 2008, Lake Vermilion was the municipal water supply for a population of 61,500 spread across the City of Danville, four nearby villages, and much of the surrounding rural area. \\

\noindent \textbf{Stream Flow Model-Data Collection }
To recover a SINDy model of water supply to Lake Vermilion, we use historical climate and streamflow data for the North Fork Vermilion watershed and river respectively. Climatic factors of specific humidity, precipitation, downwelling solar radiation, minimum and maximum air temperature, vapor pressure deficit, and wind speed near surface for each day between 1950 to 2005 were averaged over the watershed area above the streamflow sampling station located near Bismark, Illinois. This data was obtained from \cite{abatzoglou2012comparison}. Streamflow data from the Bismark station is available from November 3, 1988 to September 30, 2010, in 15 minute intervals with brief periods (no longer than a week) of missing data \cite{usgs}. Since the SINDy algorithm requires a numerical time-derivative of state variable data, we use linear interpolation to fill all missing time values. To match the resolution of the climate data, the streamflow data is summed to daily values. The resulting time period, for which data is available for both the climatic factors as well as streamflow, ranges from 1988 until 2005 resulting in 5589 data points for use in either training or testing.  

\section{Results}

\subsection{Soybean-Diesel Plant Sparsity Adjustment}

\begin{figure}
\includegraphics[width=14 cm]{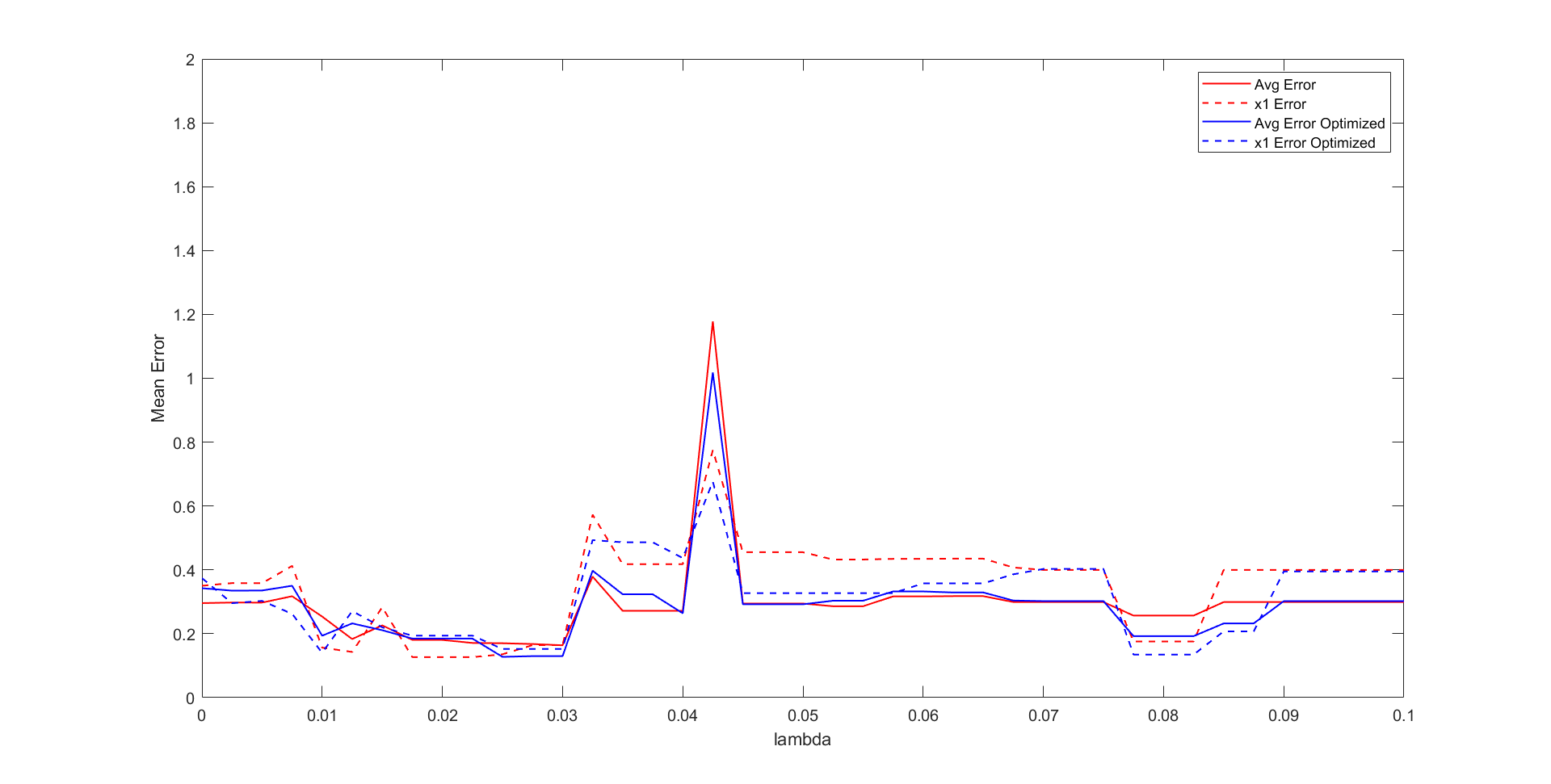}
\caption{ Minimum mean error over each set of five validation folds for $0 \leq \lambda \leq 0.1$.
\label{fig:mean_error_lambda}}
\end{figure}

We explore a range of thresholding values to vary the sparsity of the coefficient matrix for models with a function library including only first order functions. As a metric of comparison between models, we use  MAE averaged over all five validation folds for each model. We first explore the model performance for $0 \leq \lambda \leq 0.1$ with a resolution of 0.0025. Values of $\lambda$ around $0.1$ result in an overly sparse matrix with all terms equal to zero, while $\lambda = 0$ results in no forced sparsity. The resulting plot of mean error between all six state variables as well as for $x_1$ individually is shown in Figure \ref{fig:mean_error_lambda} for both the SINDy derived models as well as those further optimized. We see in Figure \ref{fig:mean_error_lambda} that the optimized models do not necessarily result in lower error versus the standard SINDy models when compared against the validation data. 

In Figure \ref{fig:mean_error_lambda}, we see minimized error between all six state variables at $\lambda = 0.025$, and minimized error for $x_1$ at $\lambda = 0.01$ and $\lambda = 0.08$. The ranges of $\lambda$ around these values are further explored with a resolution of $0.001$ to verify that error cannot be reduced further, however, adjacent values of $\lambda$ with higher performing models were not found.

\subsection{Soybean-Diesel Model Performance}

\subsubsection{5-Fold Validation}

The MAE for each of the five validation folds is shown in Table \ref{tab:5_fold_val} for the three models with sparsity parameters of $\lambda=0.01,0.025,0.08$. We look at the model results for the average of the six state variables as well as $x_1$ specifically, since accurate modeling of the output is a priority. Looking at the MAE values for each fold, we see that the model with the lowest error for both $x_1$ as well as the average of all variables varies between the standard SINDy derived model and the optimized model. This matches what can be seen in Figure \ref{fig:mean_error_lambda} across all models, which suggests that the added optimization is not necessarily resulting in overfitting to the training data.  Additionally, we see that validating on the first fold tends to yield the highest error, which indicates there may be certain dynamics during the initialization of the system that do not continue in normal operation. For the models with $\lambda = 0.01,0.025$, the lowest error is found in the third validation fold, while for the model with $\lambda = 0.08$, the lowest error is found in the fifth validation fold. Lowest error in the third validation fold is expected due to the symmetry of training data and the placement of the third validation fold in the middle. Overall, the lowest error is found in the third validation fold of the model with $\lambda = 0.025$. 

\begin{table}
\caption{ Error as MAE between five validation folds and integrated models. \label{tab:5_fold_val}}
\begin{tabular}{cccccccc}
\toprule
 \textbf{variable(s)}& \textbf{optimized} & $\lambda$ &\textbf{Fold 1}	& \textbf{Fold 2}& \textbf{Fold 3} & \textbf{Fold 4} & \textbf{Fold 5}\\
\midrule
avg($x_1,...,x_6$) & no & 0.01 & 1.0980 &   0.3590  &  0.2537 &   0.6823  &  0.3175\\
$x_1$ &              no & 0.01 & 0.9920  &  0.4983  &  0.1565  &  0.8518  &  0.4125\\
avg($x_1,...,x_6$) & yes & 0.01 & 0.6116  &  1.1461  &  0.1939  &  0.4164  &  0.3649\\
$x_1$ &              yes & 0.01 & 0.4408  &  1.8151  &  0.1404  &  0.3614  &  0.5263\\

avg($x_1,...,x_6$) & no & 0.025 & 1.4653  &  0.9635  &  0.1690  &  0.8549  &  0.1716\\
$x_1$ &              no & 0.025 & 1.7164  &  1.4017  &  0.1376  &  1.1080  &  0.2227\\
avg($x_1,...,x_6$) & yes & 0.025 & 0.9535  &  1.9236  &  0.1280  &  0.7845  &  0.2754\\
$x_1$ &              yes & 0.025 & 0.9513  &  2.8761  &  0.1479  &  0.9826  &  0.3526\\

avg($x_1,...,x_6$) & no & 0.08 & 1.7184  &  0.5653  &  0.4265  &  0.5408  &  0.2991\\
$x_1$ &              no & 0.08 & 2.3015  &  0.9276  &  0.7369  &  0.5225  &  0.3998\\
avg($x_1,...,x_6$) & yes & 0.08 & 1.2498  &  0.3070  &  0.4016  &  0.5083  &  0.3022\\
$x_1$ &              yes & 0.08 & 1.7074  &  0.4128  &  0.6963  &  0.3952  &  0.4149\\
\bottomrule
\end{tabular}
\end{table}

\subsubsection{Test Data}
We use the same error metric of MAE to judge the performance of the the models with $\lambda = 0.01,0.025,0.08$ on the test datasets of 50 hours and 200 hours. These results are tabulated in Table \ref{tab:test_data_table}. 

We see that for the dataset comprised of the 50 hours of simulation data following the training data, the standard SINDy model outperforms the optimized model for both the average error between all six state variables as well as the $x_1$ individually. However, for the long term test data of 200 hours, the optimized model performs significantly better. For the 50 hours of test data, the standard SINDy model with $\lambda = 0.01$ results in the lowest error. However, for the long term dataset of 200 hours, the optimized model with $\lambda = 0.025$ results in the lowest error, while the standard SINDy model with $\lambda = 0.01$ results in the highest amount of error. This seems to suggest that the standard SINDy models may be overfitting to a particular aspect of the simulation data from which both the training dataset and 50 hour dataset are taken. 

\begin{table}
\caption{ Error as MAE between both short and long sets of test data and integrated models.\label{tab:test_data_table}}
\begin{tabular}{cccccccc}
\toprule
 \textbf{variable(s)}& \textbf{optimized}& \textbf{test data} & $\lambda = 0.01$ & $\lambda = 0.025$	& $\lambda = 0.08$ \\
\midrule
avg($x_1,...,x_6$) & no & 50 hr & 0.1521 & 0.4734 & 0.2678\\
$x_1$ &              no & 50 hr & 0.1672 & 0.6620 & 0.3921\\
avg($x_1,...,x_6$) & yes & 50 hr& 0.6038 & 0.5370 & 0.2945\\
$x_1$ &              yes & 50 hr& 0.8847 & 0.7570 & 0.4422\\

avg($x_1,...,x_6$) & no & 200 hr  & 5.3239 & 2.1925 & 1.3440\\
$x_1$ &              no & 200 hr  & 8.5559 & 3.3656 & 0.7230\\
avg($x_1,...,x_6$) & yes & 200 hr & 0.5255 & 0.2522 & 1.2656\\
$x_1$ &              yes & 200 hr & 0.7524 & 0.2948 & 0.5924\\

\bottomrule
\end{tabular}
\end{table}

\begin{figure}
\includegraphics[width=15 cm]{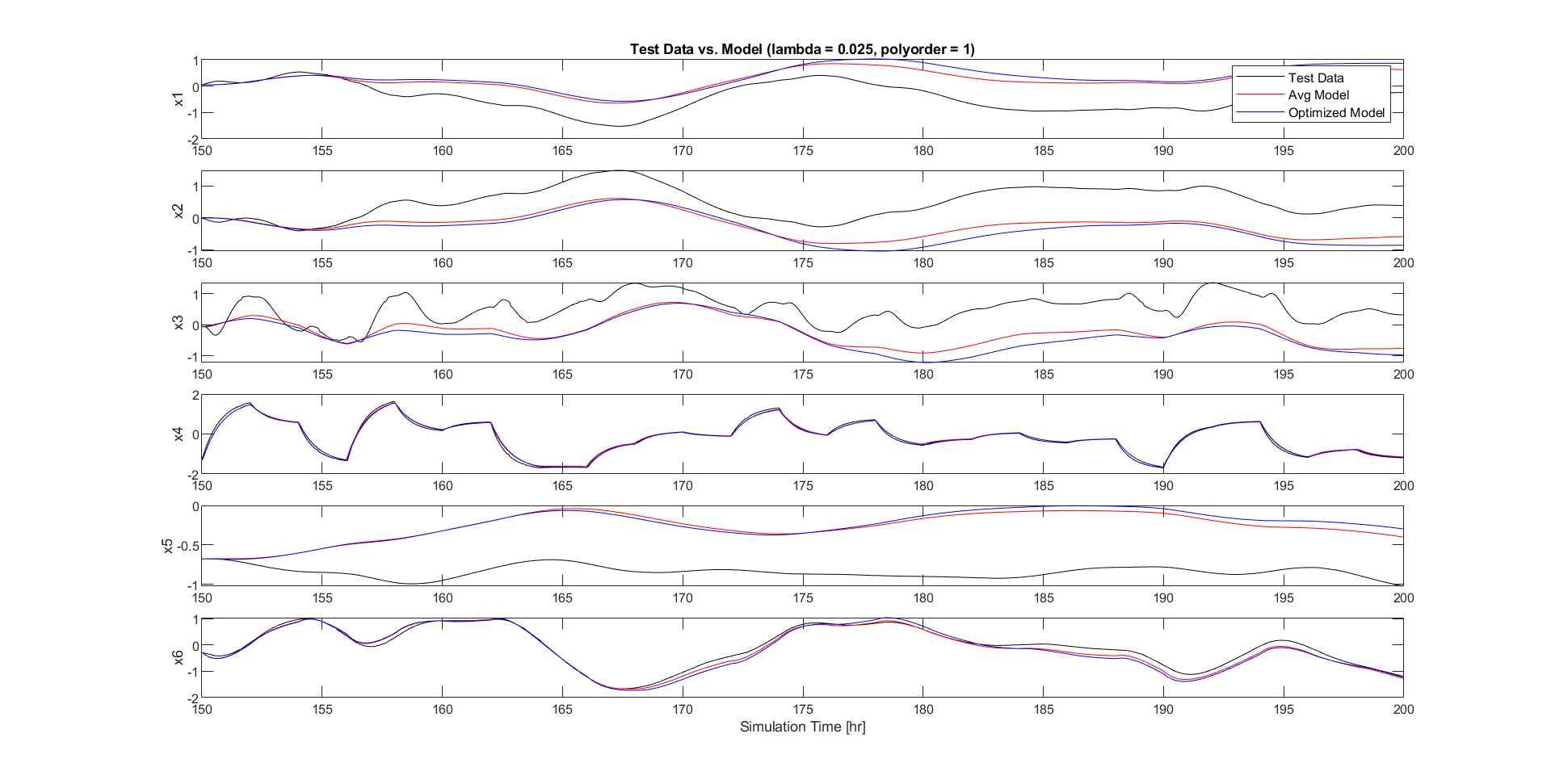}
\caption{ Integrated model results over 50 hours of test data for $\lambda=0.025$.
\label{fig:shorttest_025}}
\end{figure}

\begin{figure}
\includegraphics[width=15 cm]{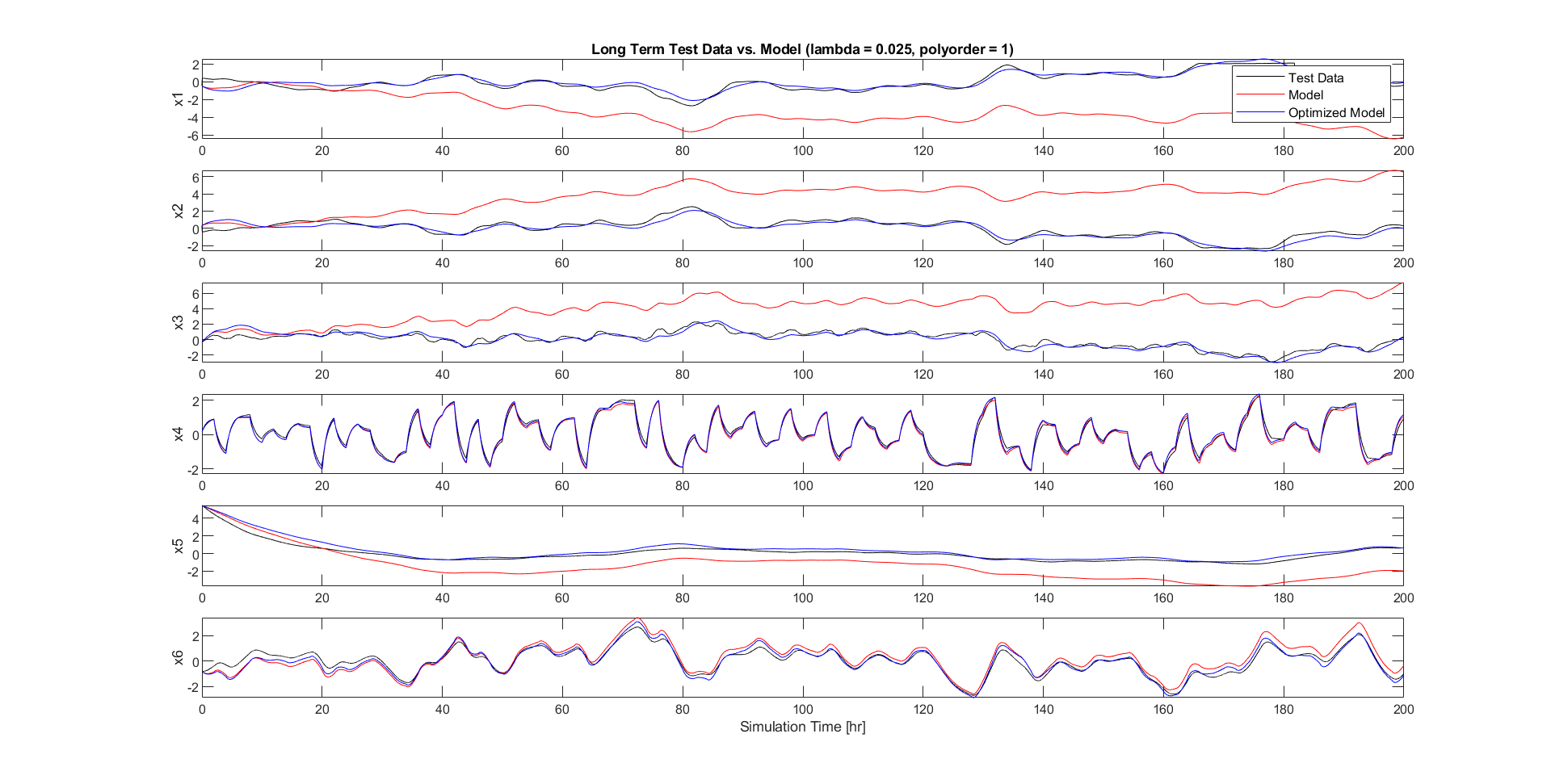}
\caption{ Integrated model results over 200 hours of test data for $\lambda=0.025$.
\label{fig:longtest_025}}
\end{figure}

Both the standard SINDy derived model with $\lambda = 0.025$, as well as the further optimized versions are integrated over time using the 50 hour and 200 hour test datasets. The resulting plots can be seen in Figures \ref{fig:shorttest_025} and \ref{fig:longtest_025} respectively. In Figure \ref{fig:shorttest_025}, we see that both the standard and optimized models fit $x_4$ and $x_6$ very well. For $x_1$, $x_2$, and $x_3$ in Figure \ref{fig:shorttest_025}, the model captures the lower frequency oscillations, but fails to reconstruct higher frequency changes. The model recreation of these state variable dynamics over time also appears to be slowly diverging from the test data, likely due to accumulation of error in the numerical integration. The worst model performance is clearly seen in the model reconstruction of $x_5$. Interestingly, across all state variables the reconstruction of the standard SINDy model is closer to the 50 hour test data than the reconstruction using the optimized model, which further suggests some type of overfitting to a particular aspect of this particular set of simulation data. 

In Figure \ref{fig:longtest_025}, we see that in the case of all variables but $x_4$ and $x_6$, the standard SINDy model is quickly diverging from the 200 hour test data. However, unlike in Figure \ref{fig:shorttest_025}, this divergent behavior is not seen in the optimized model, which while still failing to capture all high frequency oscillations in $x_1$, $x_2$, and $x_3$, is a much closer reconstruction. Additionally, the reconstruction of $x_5$ using the optimized model in Figure \ref{fig:longtest_025} is accurate.

\subsection{Soybean-Diesel Model Structure}

To compare the structure of terms between models, we use the optimized model from each of the lowest error validation folds in Table \ref{tab:5_fold_val}. Since the optimization is only applied to nonzero terms and is bounded above and below, the structure of these state equations is identical to the standard SINDy recovered equations with only some changes to parameter values. For $\lambda = 0.01$ we have the model from the third validation fold of

\begin{equation}
    \begin{array}{l} \dot{x}_1 = 0.0243\,u_{2}-1.31\,x_{1}-1.72\,x_{2}+0.451\,x_{3}-0.0344\,x_{4}-0.0696\,x_{5}-0.0289\,x_{6}\\ \dot{x}_2 = 1.27\,x_{1}-0.0273\,u_{2}-0.00935\,u_{1}+1.66\,x_{2}-0.431\,x_{3}+0.0215\,x_{4}+0.0735\,x_{5}\\+0.0282\,x_{6}\\ \dot{x}_3 = 1.09\,x_{1}-0.0332\,u_{2}-0.0738\,u_{1}+1.72\,x_{2}-0.605\,x_{3}+0.151\,x_{4}+0.0844\,x_{5}-0.123\,x_{6}\\+0.027\\ \dot{x}_4 = 0.0117\,u_{2}-1.73\,u_{1}+0.0626\,x_{2}-0.0823\,x_{3}-1.3\,x_{4}+0.0192\,x_{6}+0.00423\\ \dot{x}_5 = 0.0385\,x_{6}-0.534\,x_{2}-0.00384\,x_{3}-0.0693\,x_{5}-0.544\,x_{1}-0.0233\\ \dot{x}_6 = 0.174\,u_{1}-0.0055\,u_{2}-0.607\,x_{1}-0.945\,x_{2}+0.354\,x_{3}+0.472\,x_{4}-0.0775\,x_{5}-0.135\,x_{6} \end{array}
\label{eq:01}
\end{equation}

\noindent for $\lambda = 0.025$ the third validation fold model is given by

\begin{equation}
    \begin{array}{l} \dot{x}_1 = 0.0317\,u_{2}-1.16\,x_{1}-1.57\,x_{2}+0.436\,x_{3}-0.046\,x_{4}-0.0685\,x_{5}-0.0218\,x_{6}\\ \dot{x}_2 = 1.12\,x_{1}-0.0331\,u_{2}+1.52\,x_{2}-0.424\,x_{3}+0.0455\,x_{4}+0.0702\,x_{5}+0.0171\,x_{6}\\ \dot{x}_3 = 0.885\,x_{1}-0.0886\,u_{1}+1.56\,x_{2}-0.659\,x_{3}+0.138\,x_{4}+0.0737\,x_{5}-0.124\,x_{6}\\ \dot{x}_4 = 0.138\,x_{2}-1.63\,u_{1}-0.142\,x_{3}-1.18\,x_{4}\\ \dot{x}_5 = 0.0367\,x_{6}-0.441\,x_{2}-0.0457\,x_{5}-0.454\,x_{1}\\ \dot{x}_6 = 0.181\,u_{1}-0.548\,x_{1}-0.893\,x_{2}+0.36\,x_{3}+0.469\,x_{4}-0.0739\,x_{5}-0.139\,x_{6} \end{array}
\label{eq:025}
\end{equation}

\noindent and for $\lambda = 0.08$ we have the fifth validation fold model in the form of

\begin{equation}
    \begin{array}{l} \dot{x}_1 = 0.373\,x_{3}-0.729\,x_{2}-0.399\,x_{1}\\ \dot{x}_2 = 0.328\,x_{1}+0.634\,x_{2}-0.35\,x_{3}\\ \dot{x}_3 = 0.993\,x_{1}+1.77\,x_{2}-0.751\,x_{3}+0.245\,x_{4}+0.0816\,x_{5}-0.147\,x_{6}\\ \dot{x}_4 = -1.8\,u_{1}-1.35\,x_{4}\\ \dot{x}_5 = 0\\ \dot{x}_6 = 0.168\,u_{1}-0.65\,x_{1}-1.02\,x_{2}+0.381\,x_{3}+0.456\,x_{4}-0.086\,x_{5}-0.13\,x_{6} \end{array}
\label{eq:08}
\end{equation}

\noindent We see that as the sparsity parameter $\lambda$ increases, the sparsity of the model increases as expected with terms dropping from the model with $\lambda = 0.01$ to $\lambda = 0.08$. Several of these terms include the water input $u_2$, which falls out of $\dot{x}_4$ and $\dot{x}_6$ between Equation \ref{eq:01} and \ref{eq:025}. With $\lambda = 0.08$, terms including $u_2$ in any of the state equations are gone. Additionally, with $\lambda = 0.08$, the equation for $\dot{x}_5$ is overly sparse with only a constant zero term. Ultimately, as the sparsity is increased all state equations will go to zero. 

As terms fall out due to the increased sparsity, some remaining terms retain coefficients in the same vicinity as the previous equations, while others change significantly. For example, $\dot{x}_4$ loses all terms except for $u_1$ and $x_4$, but keeps similar coefficients, while $\dot{x}_1$ loses all terms except for $x_1$, $x_2$, and $x_3$, of which $x_1$ and $x_2$ are significantly different. 

From these equations, we can see that a basic mass balance is captured along with other expected linear behavior. In Equation \ref{eq:025}, we see that the rate of change of $x_1$, the soybean diesel output, is impacted positively by $x_3$, the input to the Diesel RadFrac column in Figure \ref{fig:process_model}, and $u_2$, the water input driving the separation process in the WashCol in Figure \ref{fig:process_model}. Likewise, this rate of change is negatively impacted by $x_2$ and $x_5$, both additional outputs from the Diesel RadFrac distillation column. The recycling stream from the MeOH RadFrac distillation column, $x_6$, also has a negative impact on the rate of change in $x_1$. Perhaps surprisingly, $x_1$ has a large negative impact on its own rate of change. This is likely due to the dynamics of saturation in the distillation column (i.e. a larger value of $x_1$ means that in the next time step the Diesel RadFrac column cannot process as much material). The structure of the state equation for $x_2$ inversely mirrors that of $x_1$.  For the rate of change of $x_3$, the non-Glycerol output of the WashCol in Figure \ref{fig:process_model}, we see a positive impact from $x_1$, $x_2$, $x_5$, and $x_4$, and a negative impact from $x_6$, which makes intuitive sense from a mass balance view. Additionally, we see a negative impact from the soybean oil input, $u_1$, and $x_3$ itself likely due to saturation dynamics. In the state equation for $x_4$, the reactor output, we see a positive impact from $x_2$ and large negative impacts from the soybean oil input, $u_1$, as well as $x_3$ and $x_4$ also likely capturing the dynamics of over saturation. For the state equation for $x_5$, we see a negative impact from $x_1$ and $x_2$, reflecting the mass balance at the Diesel RadFrac column, and a small negative impact from $x_5$ itself, likely capturing the effect of over saturation. Lastly, for the recycled flow stream, $x_6$, emerging from the MeOH RadFrac column, we see a strong positive impact from $x_4$ and $u_1$, which makes sense from a mass balance perspective. Interestingly, there is an additional positive impact from $x_3$, while $x_1$, $x_2$, and $x_5$ all provide a negative impact on the rate of change of $x_6$, which does not fit into a mass balance understanding. These terms and approximate parameter values persist even when the sparsity is somewhat increased, as evidenced in Equation \ref{eq:08}.

\subsection{Stream Flow Model Sparsity Adjustment}

\begin{figure}
\includegraphics[width=14 cm]{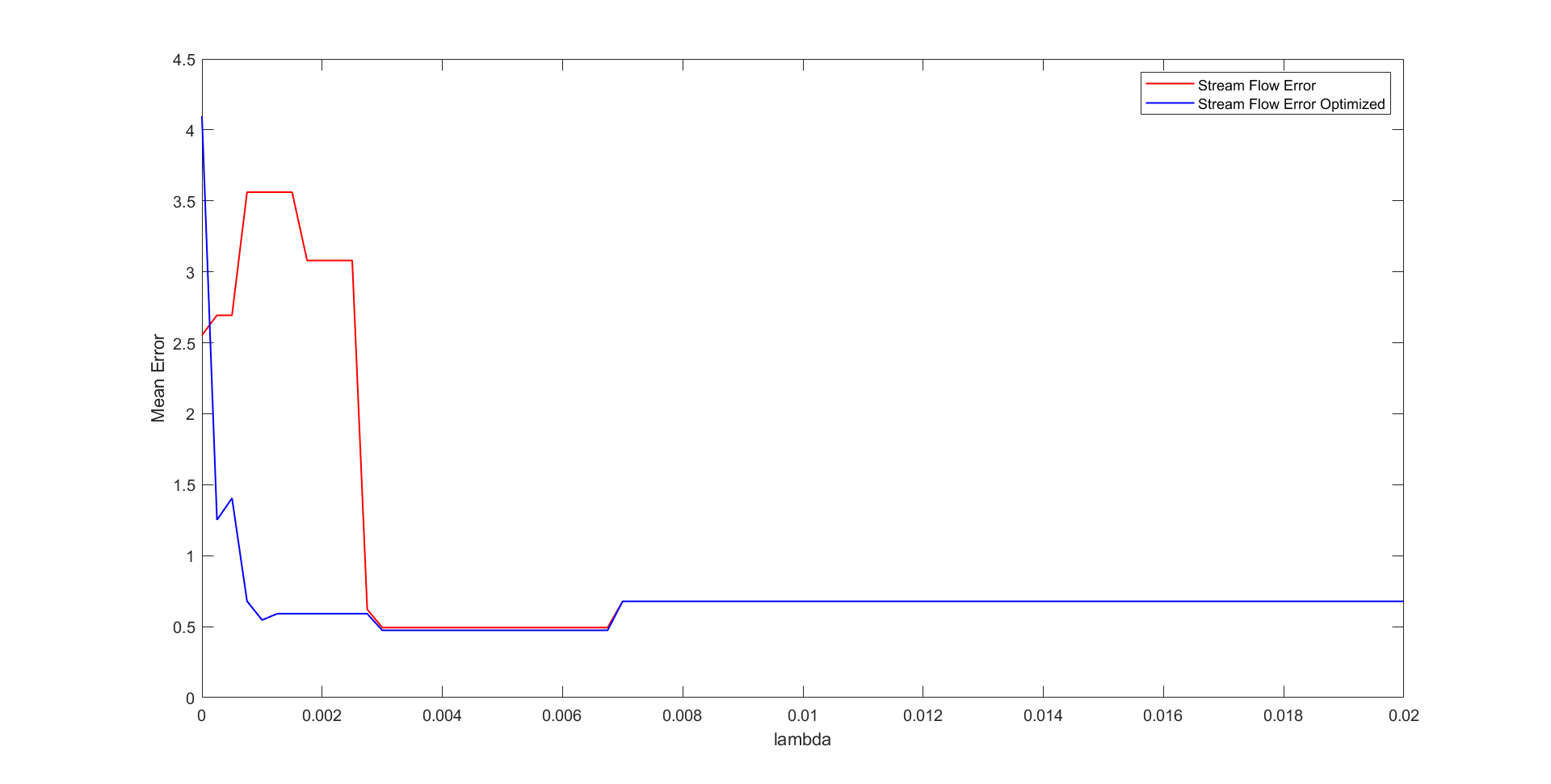}
\caption{ Minimum mean error over each set of five validation folds for $0 \leq \lambda \leq 0.02$.
\label{fig:mean_error_lambda_stream_1st}}
\end{figure}

\begin{figure}
\includegraphics[width=14 cm]{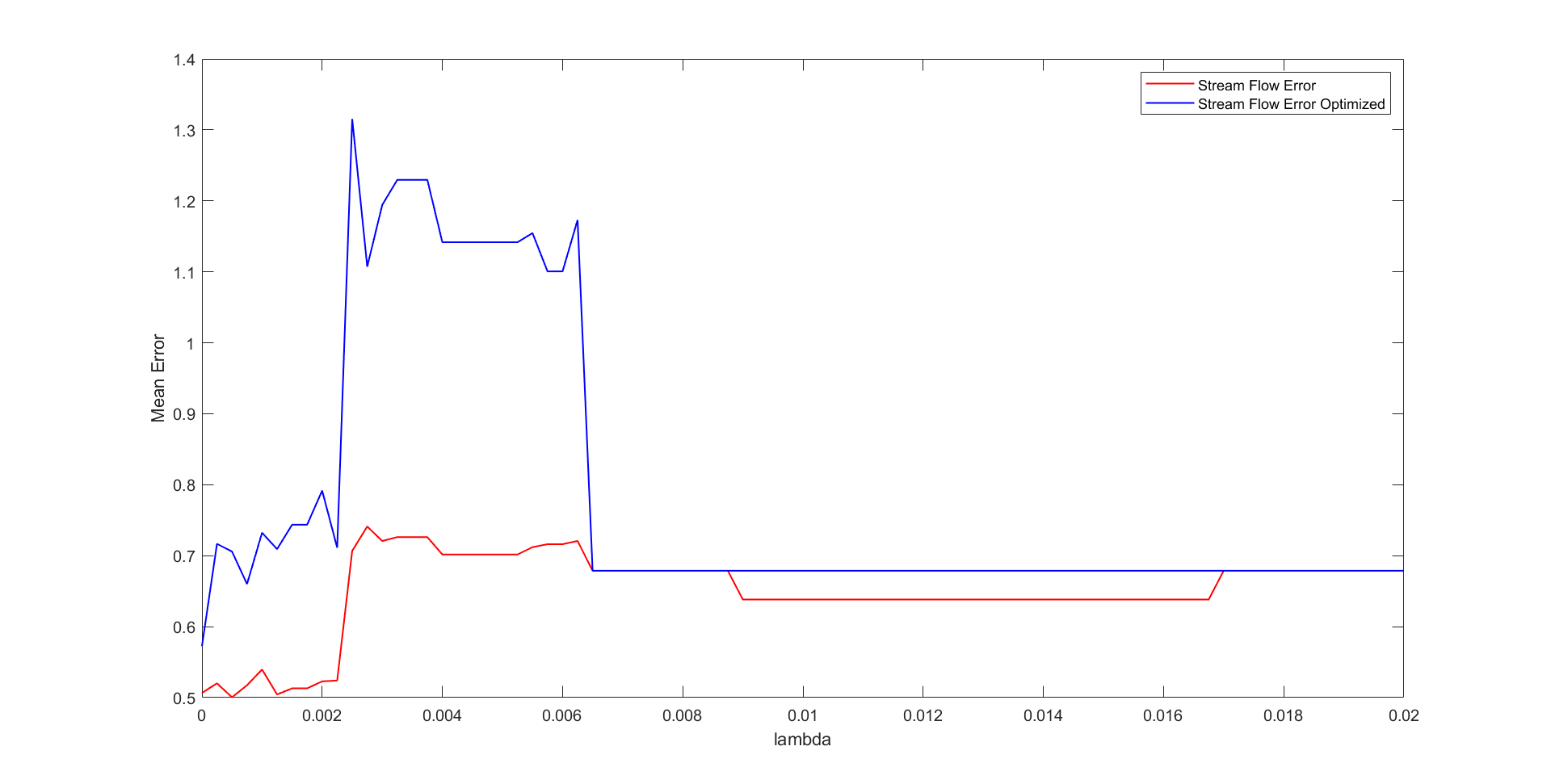}
\caption{ Minimum mean error over each set of five validation folds for $0 \leq \lambda \leq 0.02$.
\label{fig:mean_error_lambda_stream_2nd}}
\end{figure}

\begin{figure}
\includegraphics[width=14 cm]{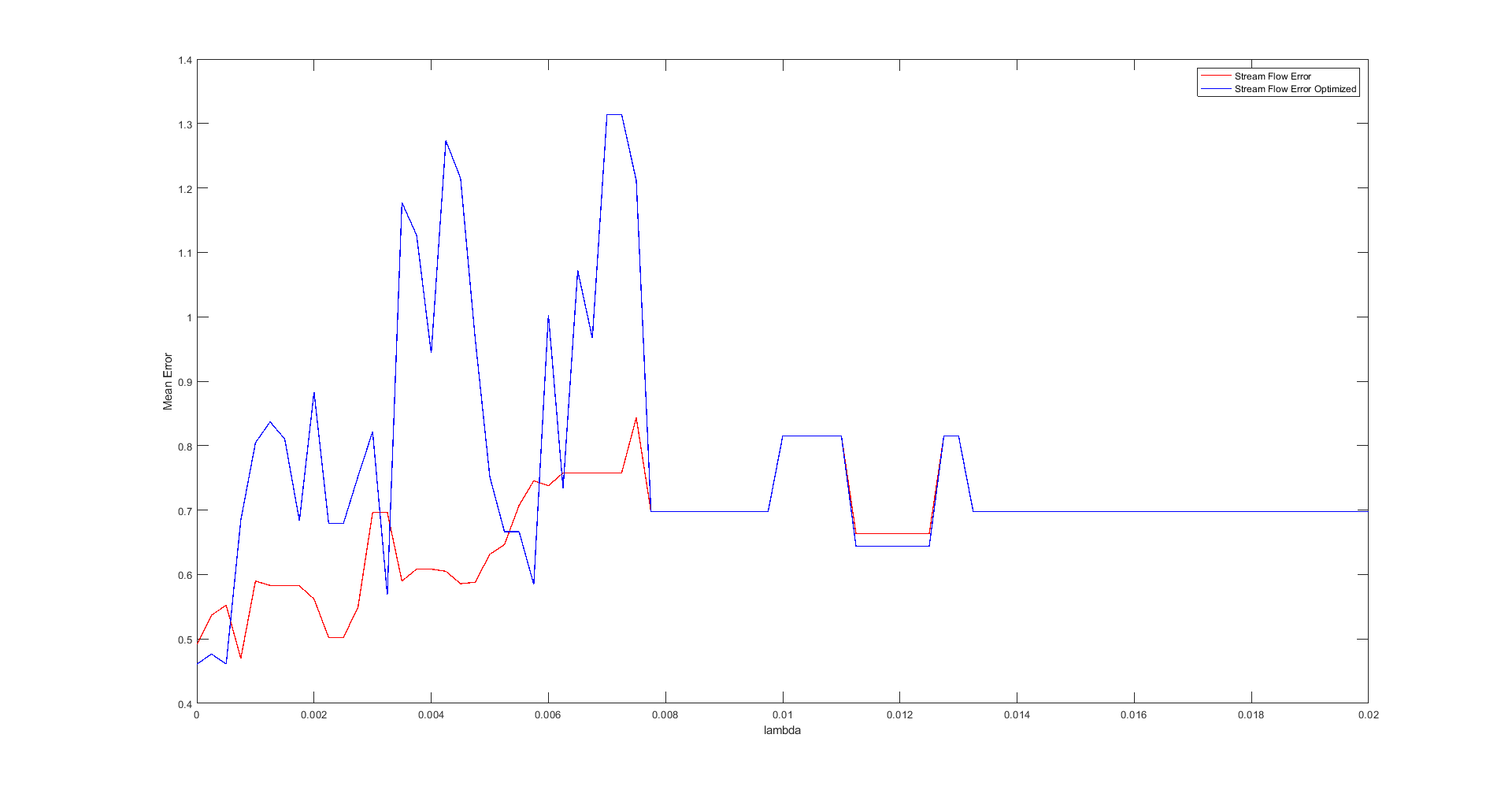}
\caption{ Minimum mean error over each set of five validation folds for $0 \leq \lambda \leq 0.02$.
\label{fig:mean_error_lambda_stream_2nd_hyst}}
\end{figure}

We again explore the range of thresholding values for which the model sparsity will change. However, unlike the soybean-diesel plant models discussed previously, we also consider function libraries that include second order polynomials in the function library supplied to the SINDy algorithm. Additionally, we consider models for which the input derivatives for which the input derivatives are included as inputs themselves. The accuracy metric of MAE is averaged over all five validation folds for each standard SINDy and further optimized model. We explore model performance over $0 \leq \lambda \leq 0.02$ for both the linear and nonlinear function libraries as seen in Figures \ref{fig:mean_error_lambda_stream_1st}, \ref{fig:mean_error_lambda_stream_2nd}, and \ref{fig:mean_error_lambda_stream_2nd_hyst}.

From Figure \ref{fig:mean_error_lambda_stream_1st} we choose to further examine $\lambda=0.006$, as there is no change in performance in the surrounding space. From Figure \ref{fig:mean_error_lambda_stream_2nd}, we see lowest error values for $\lambda$ at $0.0005$ and $0.00125$, however, we also choose to further examine $\lambda=0.012$ as a model in the last range of $\lambda$ before model reduction to 0. Lastly, from Figure \ref{fig:mean_error_lambda_stream_2nd_hyst}, we choose to further examine models for which $\lambda = 0.00175$ and $\lambda = 0.00325$. As seen in Figures \ref{fig:mean_error_lambda_stream_2nd} and  \ref{fig:mean_error_lambda_stream_2nd_hyst}, the further optimized models no longer result in reduced MAE for all values of $\lambda$, despite providing a better approximation of abrupt changes in streamflow than that of the standard model.

\subsection{Streamflow Model Performance}

The MAE for each of the five validation folds is shown in Table \ref{tab:5_fold_val_stream_flow}, where the sparsity parameter of $\lambda=0.006$ is considered as a linear model and $\lambda=0.0005,0.00125,0.012$ are considered for nonlinear models. Models that failed during integration, due either to stiffness or unbounded behavior, are marked with `NaN' rather than an MAE error value. 

\begin{table}
\caption{ Error as MAE between five validation folds and integrated models. \label{tab:5_fold_val_stream_flow}}
\begin{tabular}{ccccccccc}
\toprule
 \textbf{optimized} & \textbf{order}& $\dot u$  & $\lambda$ &\textbf{Fold 1}	& \textbf{Fold 2}& \textbf{Fold 3} & \textbf{Fold 4} & \textbf{Fold 5}\\
\midrule
          no & 1& no & 0.006 & 0.6978  &  0.9285  &  0.6334  &  1.0732  &  0.8815 \\

        yes & 1& no & 0.006 & 0.6978 &   1.1298 &   0.6416  &  0.8998 &   1.2655   \\

          no & 2& no & 0.0005 &  0.5171  &     NaN  &  0.7712  &     NaN  &  1.1387 \\

        yes & 2& no & 0.0005 &  0.4701   &    NaN  &  0.8118   &    NaN  &  1.2165 \\

         no & 2& no & 0.00125 &  1.0376  &  1.2995  &  0.9083  &  1.2114  &  0.9594  \\

       yes & 2& no & 0.00125 &  0.6302  &  5.1275  &  0.7418 &  16.9459  &  1.2391 \\

        no & 2& no & 0.012 &  0.6978  &  8.2418  &  9.8995 &   1.4322  &  6.0414  \\

         yes & 2& no & 0.012 &   0.6978  &  0.9560  &  0.7268  &  1.4322  &  1.3760 \\
         
         no & 2& yes & 0.00175 &  0.5825  &  0.7522 &   0.6061  &  6.8310  &  1.0880 \\

         yes & 2& yes & 0.00175 &  0.9897 &   3.1199  &  0.6834  & 27.9328 & NaN \\
         
         no & 2& yes & 0.00325 &  0.8024 &   0.7121  &  0.6966  &  0.8955  &  1.1799 \\

       yes & 2& yes & 0.00325 &  0.5688 &   4.8437  &  0.8218  &  1.3573  &  1.5917 \\

\bottomrule
\end{tabular}
\end{table}

\begin{figure}
\includegraphics[width=15 cm]{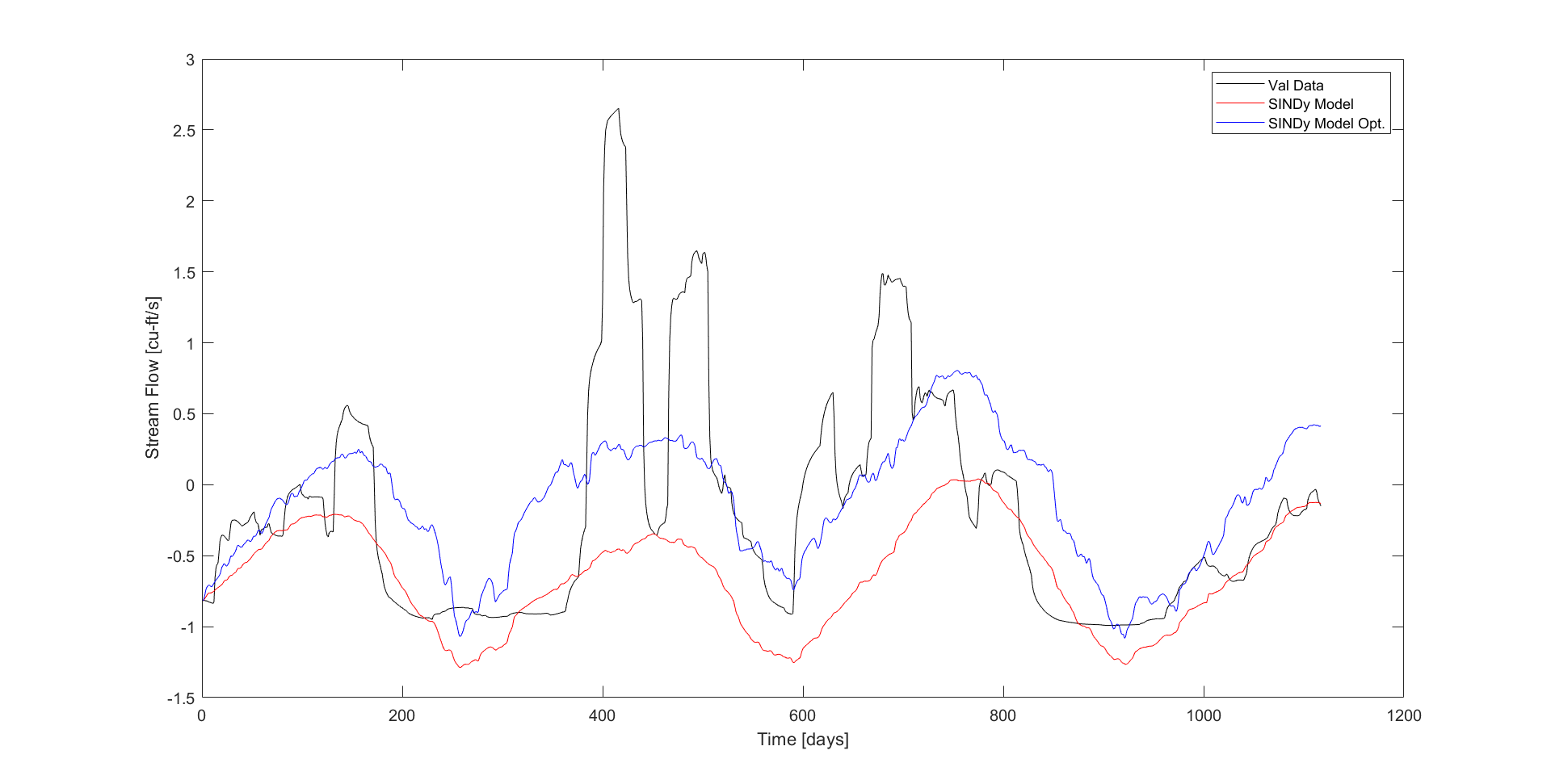}
\caption{ Nonlinear integrated model results over first validation fold for $\lambda=0.0005$.
\label{fig:0005}}
\end{figure}

\begin{figure}
\includegraphics[width=15 cm]{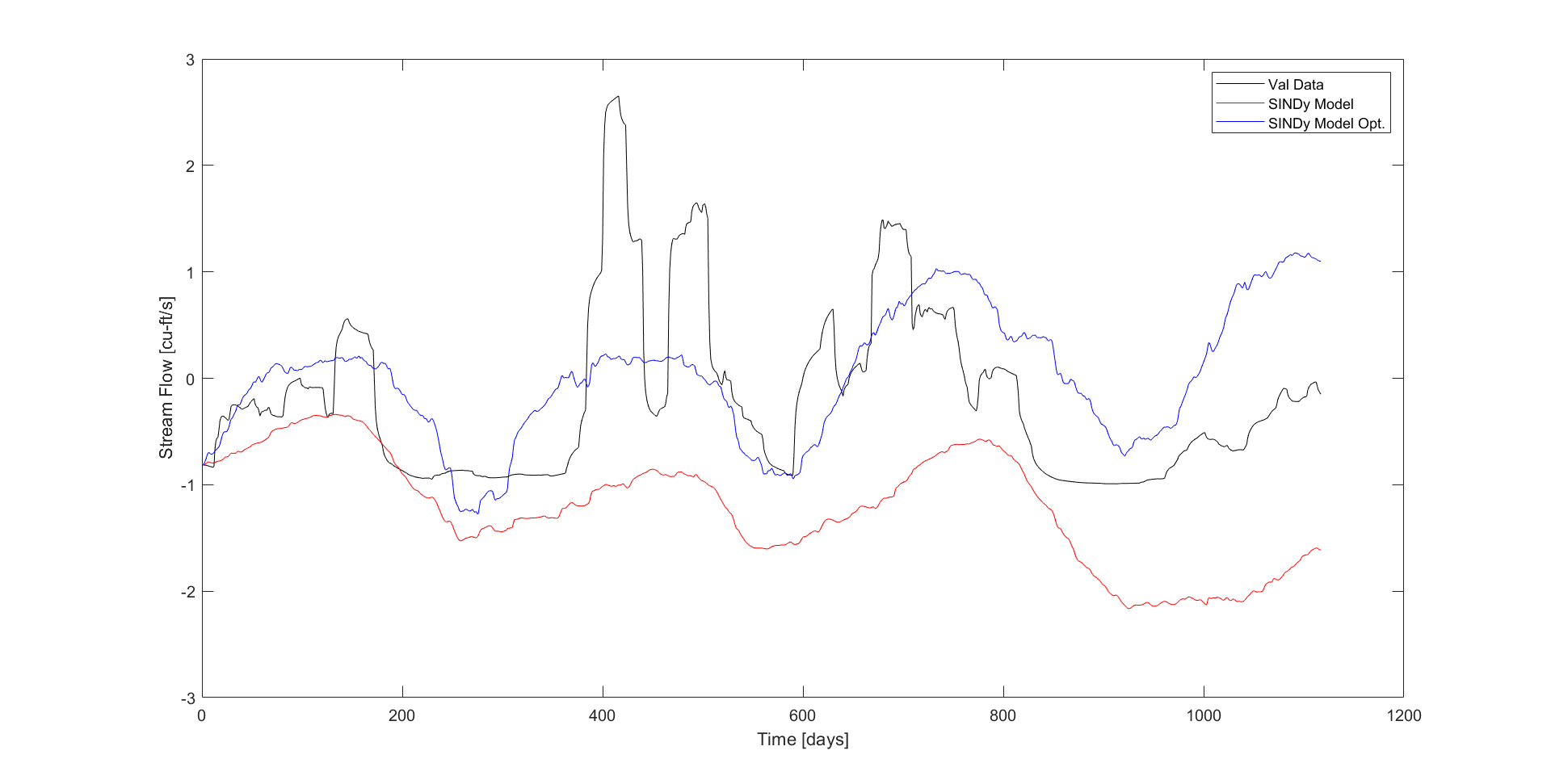}
\caption{ Nonlinear integrated model results over first validation fold for $\lambda=0.00125$.
\label{fig:00125}}
\end{figure}

In Figures \ref{fig:0005} and \ref{fig:00125} we plot the results of the integrated optimized nonlinear models for $\lambda=0.0005$ and $\lambda=0.00125$ respectively. We see that both models appear to capture seasonal streamflow tendencies, yet fail to respond to more abrupt changes happening over weeks or months.  

By contrast, in Figures {fig:00175} and {fig:00375} of integrated models containing input derivative terms, we see much better model recognition of individual streamflow peaks within a season of increased streamflow. However, these models still fail to reach the upper ranges of streamflow values and also fail to remain at low values for periods where streamflow is minimal. 
\begin{figure}
\includegraphics[width=15 cm]{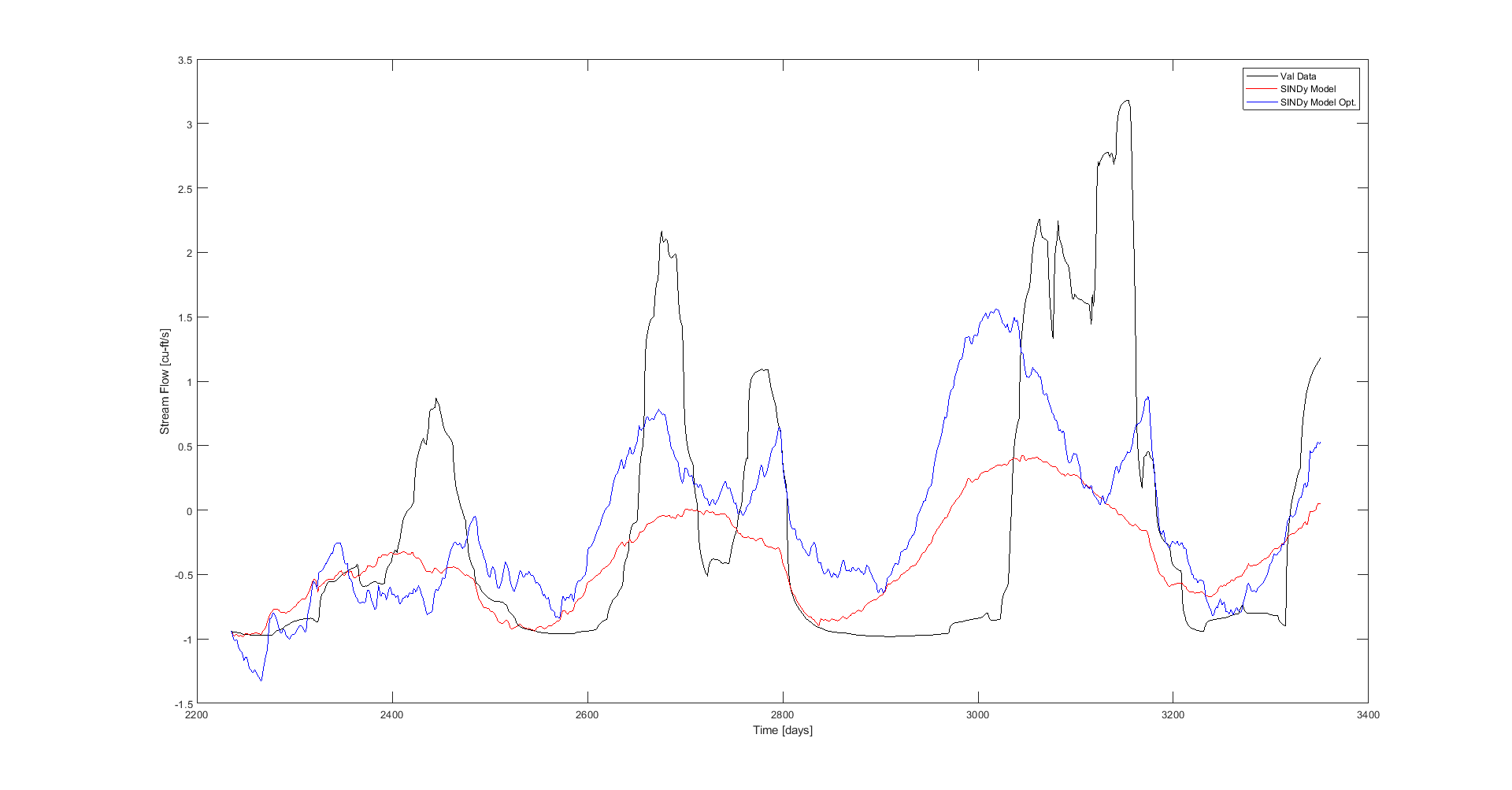}
\caption{ Nonlinear with input derivative integrated model results over first validation fold for $\lambda=0.00175$.
\label{fig:00175}}
\end{figure}

\begin{figure}
\includegraphics[width=15 cm]{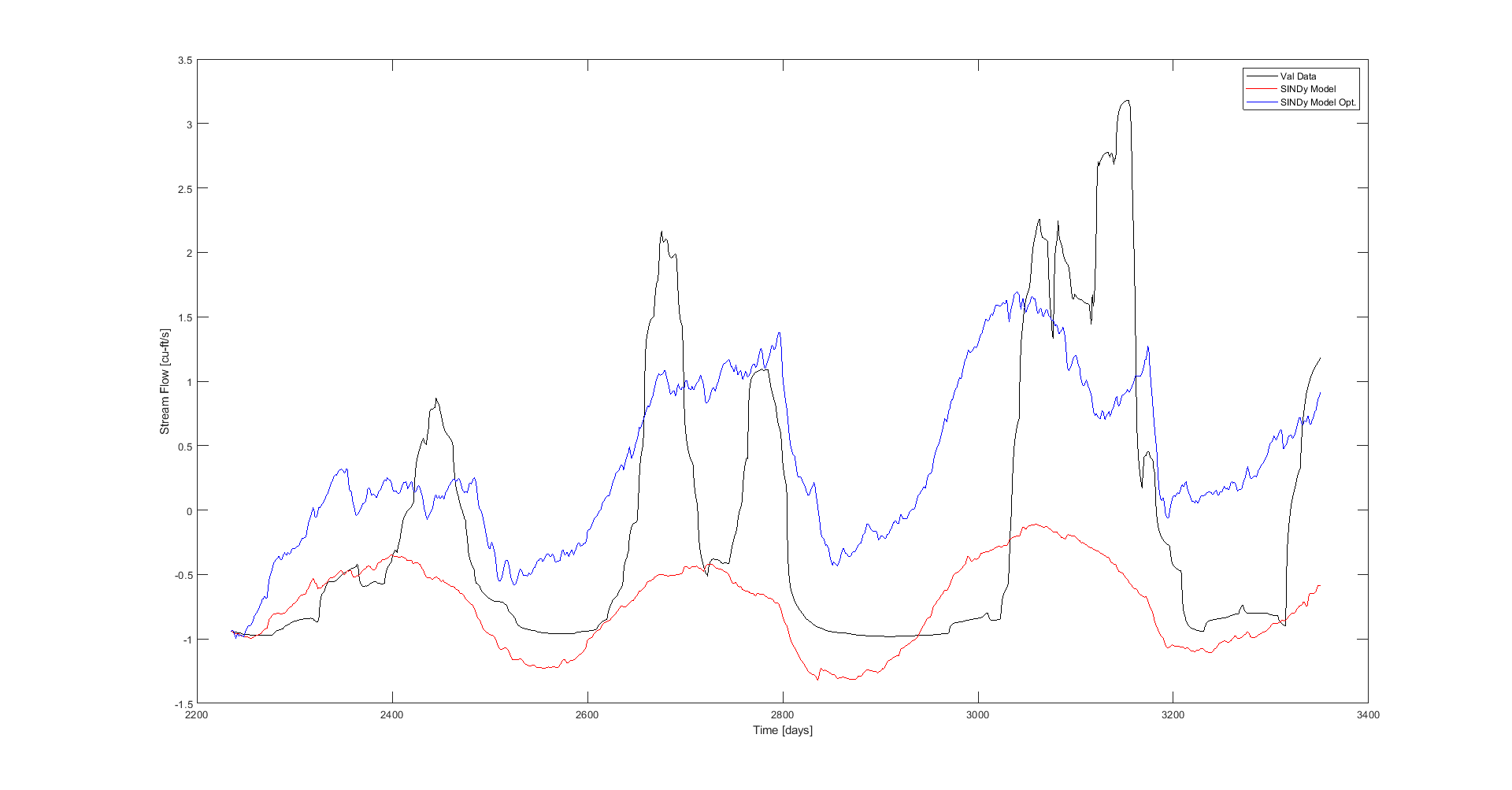}
\caption{Nonlinear with input derivative integrated model results over first validation fold for $\lambda=0.00175$.
\label{fig:00375}}
\end{figure}

\subsection{Streamflow Model Structure}

We compare the structure of terms between the optimized nonlinear models for $\lambda=0.00125$, the standard SINDy model, and $\lambda = 0.00175$, the model containing input derivative terms. These equations are given by Equation \ref{eq:00125} and Equation \ref{eq:00175} respectively, where $Q_{stream}$, $R_{solar}$, $P$, $T_{max}$, $T_{min}$, and $V$ refer to the river flow rate, solar radiation, precipitation, maximum daily temperature, minimum daily temperature, and vapor pressure deficit.

\begin{equation}
    \begin{array}{l} \dot{Q}_{stream} = 
    0.00781\,P-0.024\,T_{{min}}+0.0191\,T_{{max}}-0.0112\,V-0.0138\,P\,Q_{{stream}}\\
    -0.0048\,P\,R_{{solar}}-0.00284\,Q_{{stream}}\,R_{{solar}}+0.00352\,P\,T_{{min}}\\
    -0.0144\,P\,T_{{max}}+0.0276\,Q_{{stream}}\,T_{{min}}-0.0248\,Q_{{stream}}\,T_{{max}}\\
    +0.0157\,P\,V+0.009\,R_{{solar}}\,T_{{min}}+0.00673\,R_{{solar}}\,T_{{max}}\\
    -2.71e-4\,R_{{solar}}\,V+0.00329\,T_{{min}}\,T_{{max}}+0.0223\,T_{{max}}\,V\\
    -0.0159\,{T_{{min}}}^2-0.00754\,{T_{{max}}}^2-0.00347\,V^2-7.85e-4 \end{array}
\label{eq:00125}
\end{equation}
\noindent We see terms present in Equation \ref{eq:00125} repeated again in Equation \ref{eq:00175} with different magnitudes but the same positive or negative impact on $\dot Q_{stream}$. For example the term $-0.00754 T_{max}^2$ present in Equation \ref{eq:00125} is also present as $-0.0158 T_{max}^2$ in Equation \ref{eq:00175}. The fact that these terms remain even with consideration of additional inputs, suggests that, while unknown, they do have some relevant physical interpretation and are not the result of overfitting to noise in the streamflow or climate data. 

In Equation  \ref{eq:00175} we see that for many terms the positive or negative impact is switched when one of the component input variables is switched with its derivative. For example, the positive $R_{solar}V$ term becomes negative when either $\dot R_{solar}$ or $\dot V$ is substituted for $R_{solar}$ or $V$ respectively, but remains positive when both are substituted. This makes sense if we consider the effect of past system inputs on current system state. A positive value for either $\dot R_{solar}$ or $\dot V$ implies that the current value is larger than the previous input value and this previously smaller value exerts a negative impact on the rate of change of streamflow. Conversely, a negative value for either $\dot R_{solar}$ or $\dot V$ implies that the current value is smaller and this past large value exerts a positive impact on  $\dot Q_{stream}$. The number of input derivative terms appearing in the model structure suggests a heavy dependence on input history and additional steps to address this hysteresis are likely necessary to improve the accuracy of the model. 
\vspace{-0.10in}
\begin{equation}
\begin{array}{l} \dot{Q}_{stream} =
0.00619\,P-0.00656\,R_{\mathrm{solar}}+0.0167\,\dot R_{{\mathrm{solar}}}\\
-0.0475\,T_{\mathrm{min}}+0.0707\,T_{\mathrm{max}}+0.0121\,\dot T_{{\mathrm{min}}}\\
-0.0141\,\dot T_{{\mathrm{max}}}-0.0208\,V-0.00707\,\dot{V}-0.0235\,P\,Q_{\mathrm{stream}}\\
+0.0133\,\dot{P}\,Q_{\mathrm{stream}}-0.00506\,Q_{\mathrm{stream}}\,R_{\mathrm{solar}}\\
+0.00593\,Q_{\mathrm{stream}}\,\dot R_{{\mathrm{solar}}}+0.0334\,P\,T_{\mathrm{min}}\\
-0.0497\,P\,T_{\mathrm{max}}-0.0156\,P\,\dot T_{{\mathrm{min}}}\\
-0.0343\,P\,\dot T_{{\mathrm{max}}}-0.0173\,\dot{P}\,T_{\mathrm{min}}\\
+0.0314\,\dot{P}\,T_{\mathrm{max}}+0.0129\,\dot{P}\,\dot T_{{\mathrm{min}}}\\
+0.0224\,\dot{P}\,\dot T_{{\mathrm{max}}}+0.00748\,Q_{\mathrm{stream}}\,T_{\mathrm{min}}\\
-0.00738\,Q_{\mathrm{stream}}\,T_{\mathrm{max}}+0.0128\,Q_{\mathrm{stream}}\,\dot T_{{\mathrm{max}}}\\
+0.0304\,P\,V+0.00566\,P\,\dot{V}-0.0201\,\dot{P}\,V-0.0112\,R_{\mathrm{solar}}\,T_{\mathrm{min}}\\
+0.0282\,R_{\mathrm{solar}}\,T_{\mathrm{max}}+0.0253\,\dot R_{{\mathrm{solar}}}\,T_{\mathrm{min}}\\
-0.0233\,R_{\mathrm{solar}}\,\dot T_{{\mathrm{max}}}-0.0198\,\dot R_{{\mathrm{solar}}}\,T_{\mathrm{max}}\\
-0.0266\,\dot R_{{\mathrm{solar}}}\,\dot T_{{\mathrm{max}}}\\
+0.00654\,Q_{\mathrm{stream}}\,V+0.00344\,Q_{\mathrm{stream}}\,\dot{V}\\
+0.00506\,R_{\mathrm{solar}}\,V-0.0218\,\dot R_{{\mathrm{solar}}}\,V-0.0171\,R_{\mathrm{solar}}\,\dot{V}+0.0384\,\dot R_{{\mathrm{solar}}}\,\dot{V}\\
+0.0211\,T_{\mathrm{min}}\,T_{\mathrm{max}}+0.0572\,T_{\mathrm{min}}\,\dot T_{{\mathrm{min}}}\\
-0.0239\,T_{\mathrm{min}}\,\dot T_{{\mathrm{max}}}-0.0462\,T_{\mathrm{max}}\,\dot T_{{\mathrm{min}}}\\
+0.0579\,T_{\mathrm{max}}\,\dot T_{{\mathrm{max}}}-0.00872\,\dot T_{{\mathrm{min}}}\,\dot T_{{\mathrm{max}}}\\
+0.0099\,T_{\mathrm{min}}\,V+0.0289\,T_{\mathrm{max}}\,V-0.0646\,\dot T_{{\mathrm{max}}}\,V\\
+0.0163\,T_{\mathrm{min}}\,\dot{V}-0.0112\,T_{\mathrm{max}}\,\dot{V}+0.0623\,\dot T_{{\mathrm{max}}}\,\dot{V}\\
+0.0189\,V\,\dot{V}-0.00742\,{R_{\dot{\mathrm{solar}}}}^2\\
-0.0195\,{T_{\mathrm{min}}}^2-0.0158\,{T_{\mathrm{max}}}^2-0.0374\,{\dot T_{{\mathrm{max}}}}^2-0.0113\,V^2-0.0206\,{\dot{V}}^2-0.0112
\end{array}
\label{eq:00175}
\end{equation}

\section{Conclusions and Discussions}

In this paper, we propose using a recently developed system identification method based on sparse identification of model coefficients for fast recovery of low-order dynamic models of process industries and natural systems.
We utilize a hybrid mechanistic- machine learning approach by using the  simulated data for process flow obtained from high-order mechanistic models to train the model using machine learning. Similarly, for natural systems we use data from observations and climate models based which use physical principles.
We  modify the original SINDy method and find that further nonlinear optimization of sparsity matrix coefficients improves model performance and limits drift over time. This SINDy plus optimization method is able to recover an accurate low-order linear model and can likely be extended to nonlinear process models with some modifications to the forcing functions and selection process for state variables. We also demonstrate that these data driven methods for creating reduced order models for highly chaotic natural systems may not be adequate. However, dynamical understanding of natural systems is a prerequisite for meeting the sustainability goals and training dynamical models for ecological/natural systems is challenging. Hence, greater efforts are required to develop appropriate machine learning methods for creating reduced order models of complex natural systems. 

Future research would benefit from expanding the scope of state variables included in the process model to include other variables such as temperature and pressure of unit operation blocks and internal flows. Particularly for model applications involving control or observation, a more complete picture of the state space may be required. Incorporating these new state variables into the SINDy method will require greater consideration of the input excitation function frequency and amplitude, and  measurement frequency used during data collection to account for differing time-scales among heterogeneous state variables. Additionally, some physics-based \textit{a prior} knowledge of what sorts of functions will likely appear in the model structure will become more important to limit the possible function space and thus reduce the computational time required to solve the SINDy regression problem. 

One of the main hindrances to improving the performance of the streamflow model is the low number of data points available for training and testing as well as missing values in what data is available. This lack of data availability/completeness may be addressed through interpolation between data points to generate additional data for use. One such technique that appears promising for this application is the use of a cubic smoothing spline. Additional nonlinearity may also be required to produce the apparently chaotic behavior of this natural system. This might be achieved through the addition of higher order polynomials to the function library or even the heaviside function as an operator on inputs and state variables. 

These dynamical models can be used for sustainability analysis based on dynamical trajectory analysis of each system under varying scenarios of resource availability and climate change to provide insights into potential limits of resource availability.

\section*{Acknowledgments}
This work was supported in part by National Science Foundation (CBET 1805741) . We also thank Dr. Sebastian Oberst and Dr. Merten Stender for their feedback and suggestions for improvement of the models.

\bibliographystyle{unsrt}  
\bibliography{references}

\end{document}